\documentclass{iopart}
\usepackage{graphicx}
\begin{document}
\title{Point-contact spectroscopy of the heavy-fermion superconductor CePt$_{3}$Si}

\author{R Onuki$^{1}$, A Sumiyama$^{1}$, Y Oda$^{1}$, T Yasuda$^{2}$, R Settai$^{2}$ and Y \={O}nuki$^{2}$}

\address{                    
  $^{1}$ Department of Material Science, Faculty of Science, University of Hyogo, Ak\={o}-gun 678-1297, Japan}
\address{                    
  $^{2}$ Graduate School of Science, Osaka University, Toyonaka
560-0043, Japan
}
\ead{sumiyama@sci.u-hyogo.ac.jp}

\begin{abstract}
Differential resistance spectra (${\rm d}V/{\rm d}I-V$ characteristics) have been measured for point-contacts between the heavy-fermion superconductor (HFS) CePt$_{3}$Si and a normal metal. Some contacts show a peak at $V=0$ that is characteristic of HFS coexisting with a magnetic order such as UPd$_2$Al$_3$, UNi$_2$Al$_3$ and URu$_2$Si$_2$. The evolution of the peak occurs well above the antiferromagnetic transition temperature $T_{{\rm N}}\sim$ 2.2 K, so that the direct relationship with the magnetic transition is questionable. The half-width of the peak seems to reflect the crystal field splitting or the spin-wave gap as observed for the above-mentioned HFSs, possibly suggesting that some common scattering process induces the zero-bias peaks in these materials.
\end{abstract}


\submitto{\JPCM}

\maketitle

The heavy fermion superconductor (HFS) CePt$_{3}$Si has attracted much attention in recent years\cite{1,2}, since its noncentrosymmetric crystal structure and strong correlation between electrons may lead to the mixing of spin-singlet and spin-triplet superconductivity. The spin triplet superconductivity is supported experimentally by the large upper critical field\cite{1}  and the NMR Knight shift\cite{3}, while the Hebel-Slichter peak in the nuclear spin-lattice relaxation rate suggests conventional spin-singlet superconductivity\cite{4}. In addition to this duality, the temperature dependence of various physical properties such as thermal conductivity\cite{5} and magnetic penetration depth\cite{6} suggests the line node in the energy gap. A theoretical work that combines and explains all these results has been proposed\cite{7}.

Besides unconventional superconductivity, CePt$_{3}$Si is the first Ce-based heavy fermion superconductor that coexists with an antiferromagnetic order at ambient pressure. The spin structure has been revealed by neutron diffraction measurements\cite{8}. Its complex pressure-temperature phase diagram has been investigated by specific heat\cite{9} and magnetic susceptibility measurements\cite{10}, although the correlation between magnetism and superconductivity is still not clear. Since this issue directly relates to which symmetry of superconductivity is favored in this material, it will be useful to investigate how the electronic state is modified through the antiferromagnetic and superconducting transition. 

Point-contact spectroscopy, which measures the bias voltage dependence of the differential resistance of the point-contacts, has given fruitful information for both normal and superconducting states of heavy fermion materials\cite{11}. In the normal state, most materials show a minimum of ${\rm d}V/{\rm d}I$ at zero bias and it has been ascribed to the increase in local temperature due to Joule heating. The exceptions that show a zero-bias maximum are URu$_{2}$Si$_{2}$, UPd$_{2}$Al$_{3}$ and UNi$_{2}$Al$_{3}$\cite{11,12}. It should be noted that these materials show superconductivity that coexists with antiferromagnetism (UPd$_{2}$Al$_{3}$\cite{13}, UNi$_{2}$Al$_{3}$\cite{14}) or so-called "hidden order"(URu$_{2}$Si$_{2}$\cite{15}). It will be interesting to test whether such a spectral property is common to the heavy-fermion superconductors that coexist with a magnetic order.

In this paper, we investigate the point-contact spectroscopy of CePt$_{3}$Si and report that it also shows a similar structure at zero bias. Possible factors that determine the energy width of the structures are discussed also.

A single crystal of CePt$_{3}$Si was grown by the Bridgeman method. Details of the process for producing the crystal are provided in previous papers\cite{16,17}. In the present study, a rectangular piece with edges of about 3 mm was used; point contacts were made on the (100) and (001) faces. Hereafter, the contacts are denoted as $I \parallel a$ and $I \parallel c$, assuming that the preferred current direction is perpendicular to the surface. Prior to measurements these faces were polished using diamond polish down to a size of 1 $\mu$m to obtain a mirror-like surface. A Pt needle was gently pressed onto the sample with a commercial piezo-electric actuator (attocube ANPx100) to form a point contact between a superconductor and a normal metal. The position of the contacts were changed using another actuator (attocube ANPz100). These actuators move stepwise by about 100 nm at low temperatures, which enables precise control of the contact resistance and position. The point-contact apparatus was set to the mixing chamber of a dilution refrigerator and cooled down to 80 mK.

The differential resistance ${\rm d}V/{\rm d}I$ of the point contacts was measured by the four-wire method using a system that consists of a current source (Keithley 6221) and a nanovoltmeter (Keithley 2182). The combination of them provides a special pattern of current sweep and ${\rm d}V/{\rm d}I$ is numerically derived; a constant differential current ${\rm d}I$ is added to or subtracted from a staircase sweep, and both the differential voltage ${\rm d}V$ and the bias voltage $V$ are calculated from the voltage measured at each current step. The sign of the bias voltage $V$ is referred to the polarity of the needle.

\begin{figure}
\begin{center}
\includegraphics*[width=0.7\linewidth, trim=1cm 6cm 2cm 10cm, clip]{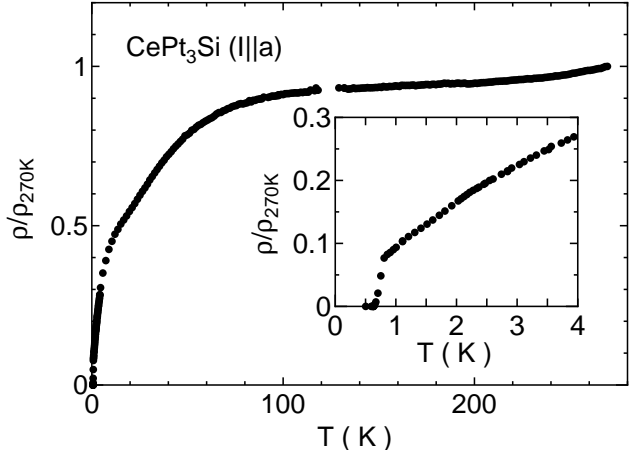}
\caption{\label{fig1}Temperature dependence of normalized resistance $\rho/\rho_{270 {\rm K}}$ of the CePt$_{3}$Si crystal used in this work. Inset shows the superconducting ($T_{{\rm c}} \sim 0.6$ K) and antiferromagnetic ($T_{{\rm N}} \sim 2.2$ K) transitions.}
\end{center}
\end{figure}
%
Figure \ref{fig1} shows the superconducting transition of the CePt$_{3}$Si crystal used for the present investigation. As the temperature is decreased, the resistance of the bulk CePt$_{3}$Si shows a gradual decrease, followed by a drop below 80 K, and then becomes zero at $T_{{\rm c}}\sim$ 0.6 K. As seen in the inset, a slight change in slope appears at $T_{{\rm N}}\sim$ 2.2 K.

\begin{figure}
\begin{center}
\includegraphics*[width=1.0\linewidth, trim=0.5cm 8cm 1cm 8cm, clip]{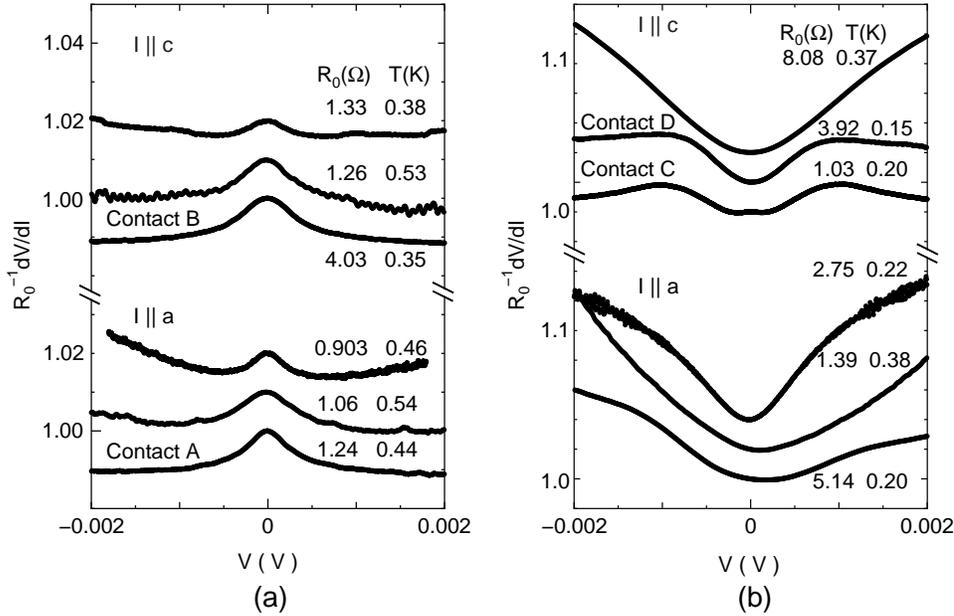}
\caption{\label{fig2}Normalized ${\rm d}V/{\rm d}I$ versus $V$ characteristics of CePt$_{3}$Si/Pt contacts below $T_{{\rm c}}$, where $R_{0}$ is the differential resistance at $V=0$. (a) and (b) are for the contacts that show a zero-bias peak and a zero-bias dip, respectively. The spectra are shifted vertically by (a) 0.01 and (b) 0.02 for clarity.}
\end{center}
\end{figure}
%
We show in figure~\ref{fig2} ${\rm d}V/{\rm d}I$ spectra taken below $T_{{\rm c}}$ for different contacts on (100) and (001) faces. The differential resistance ${\rm d}V/{\rm d}I$ for each contact is normalized by $R_{0}={\rm d}V/{\rm d}I$ at $V=0$. For both current directions, two types of spectra have been observed. One type shows a zero-bias peak with a voltage width of $\sim$1 mV (figure~\ref{fig2}(a)). The other is characterized by a minimum at zero bias with a width varying widely (figure~\ref{fig2}(b)). The lack of a clear difference between the $a$- and $c$-axis directions does not necessarily exclude the anisotropy in CePt$_{3}$Si, since the directionality tends to be lost in point-contact spectroscopy due to the roughness of the surface.
\begin{figure}
\begin{center}
\includegraphics*[width=0.8\linewidth, trim=1cm 1cm 4cm 8cm, clip]{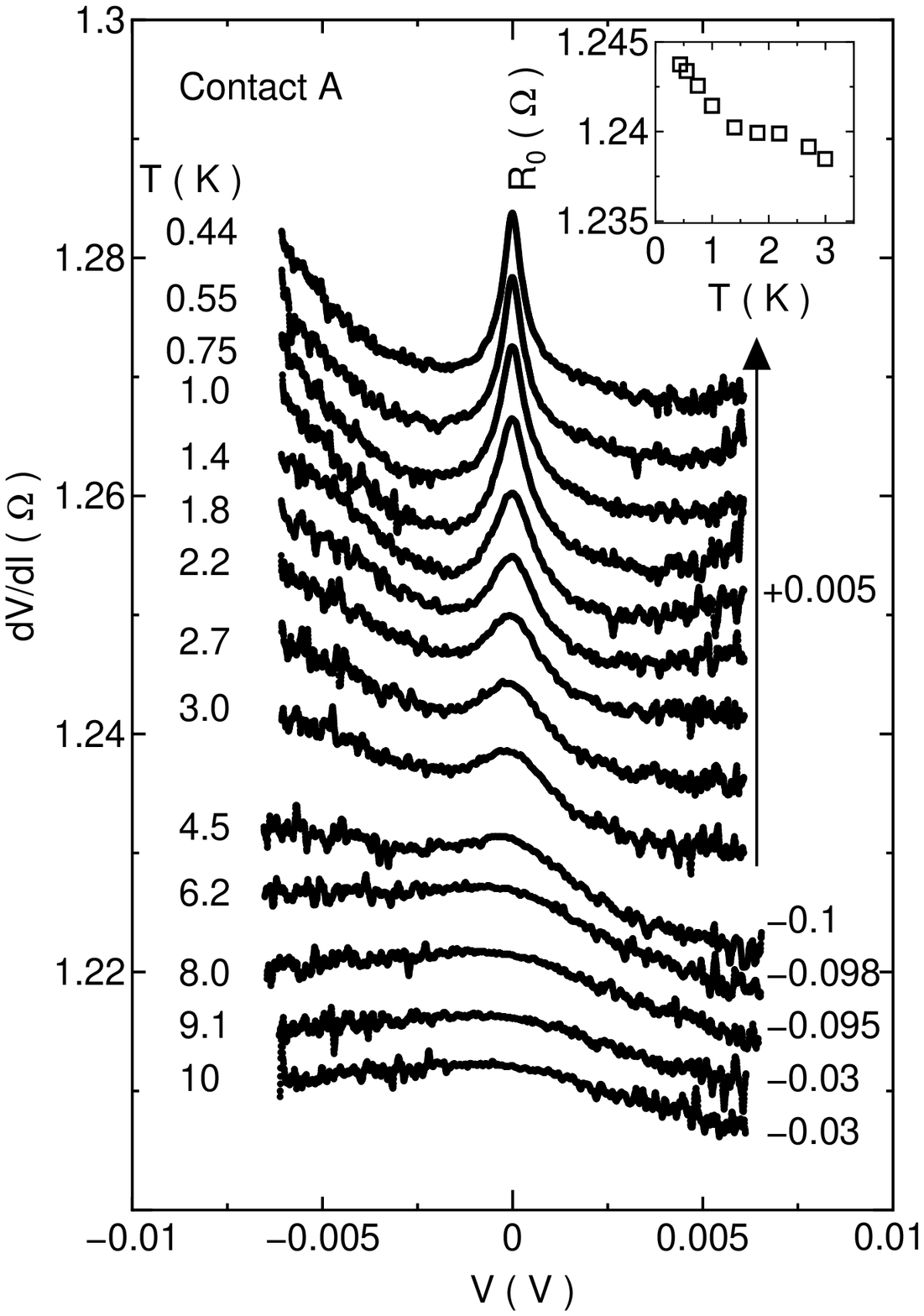}
\caption{\label{fig3}Temperature dependence of ${\rm d}V/{\rm d}I$ versus $V$ characteristics for Contact A. The spectra below 3.0 K are shifted vertically by 0.005 $\Omega$ for clarity, while the shift is indicated by the figure for those above 3.0 K. Inset: temperature dependence of $R_{0}$.}
\end{center}
\end{figure}
%

As the temperature is increased, we have observed that both types of structures gradually diminish, but still exist well above $T_{{\rm c}}$, so that they are not attributed to superconductivity. As a typical example, the temperature dependence of the spectra for Contact A in figure~\ref{fig2} (a) is shown in figure~\ref{fig3}. The temperature dependence was measured during the warming process from the lowest temperature, and we often encountered an abrupt change in $R_{0}$ above 3 K, possibly because the boiling of $^{3}$He/$^{4}$He mixture begins and causes the change in the contact condition by vibrations. In figure~\ref{fig3}, the spectra after the abrupt change in $R_{0}$ are also shown by shifting vertically by an appropriate extent. Although the curve shape also changes a little, the peak at zero bias seems to diminish successively and vanish above 4.2 K. The observation of the zero-bias peak above $T_{{\rm N}}\sim$ 2.2 K raises the question as to whether the peak structure is ascribed to the antiferromagnetism.
  
In order to clarify the origin of the peak, we should examine what kind of information the measured differential resistance contains. In the case that CePt$_{3}$Si and Pt contact through a circular plane of radius $a$, the contact resistance $R_{0}$ at zero bias is approximately expressed as the sum of Sharvin resistance $R_{{\rm SHA}}$ and Maxwell resistance $R_{{\rm MAX}}$, as given by
\begin{equation}
R\sim \frac{2R_{{\rm K}}}{(ak_{{\rm F}})^{2}}+\frac{\rho}{4a} ,
\end{equation}
where $k_{{\rm F}}$ is the Fermi wave number, $R_{{\rm K}}=h/e^{2}=25.8$ k$\Omega$, and $\rho$ is the electrical resistivity of CePt$_{3}$Si near the contact\cite{18,19}. The $R_{{\rm MAX}}$ due to Pt resistivity has been neglected, because it is supposed to be small and independent of temperature at low temperatures. The two parts become comparable at a resistance of $R_{{\rm eq}}\sim \rho^{2}k_{{\rm F}}^{2}/16R_{{\rm K}}$. If $\rho \sim 1$ $\mu \Omega \cdot $cm\cite{16} is used and a typical metallic $k_{{\rm F}} \sim$10 nm$^{-1}$ is assumed, $R_{{\rm eq}}$ is about 0.02 $\Omega $; if $R_{0} > R_{{\rm eq}}$ as seen in the present contacts, $R_{{\rm SHA}}$ is dominant and the data contain intrinsic spectroscopic information. In the actual point contacts, however, $\rho$ near the surface is possibly larger than that of the bulk, and an oxide layer that inhibits a clean metallic contact may be formed on the CePt$_{3}$Si surface. Consequently the dominant part of resistivity should be judged not only from the $R_{0}$ value but also from the behaviour of ${\rm d}V/{\rm d}I$. 

As seen in the inset of figure~\ref{fig3}, $R_{0}$ increases as the temperature is lowered. This cannot be explained by the temperature dependence of $\rho(T)$ in the Maxwell resistance, since the resistance of bulk CePt$_{3}$Si decreases by lowering the temperature, as shown in figure~\ref{fig1}. The increase in $R_{0}$ should be ascribed to $R_{{\rm SHA}}$ and contain some spectroscopic information. Such peak structures have been observed for only a few heavy-fermion systems: point-contact spectroscopy of URu$_{2}$Si$_{2}$\cite{11} and scanning tunneling spectroscopy of URu$_{2}$Si$_{2}$, UPd$_{2}$Al$_{3}$ and UNi$_{2}$Al$_{3}$\cite{12}. It should be noted that all these materials including CePt$_{3}$Si are heavy-fermion superconductors that coexist with an antiferromagnetic order, although for URu$_{2}$Si$_{2}$ the order is now regarded as some other type, which is so-called hidden order.

For most of heavy-fermion systems, a minimum in ${\rm d}V/{\rm d}I$ at zero bias is usually observed as seen in figure~\ref{fig2} (b) and is explained by a heating or thermal model; if the inelastic mean free path $\ell_{{\rm in}}$ is much smaller than the contact size $a$, $R_{{\rm MAX}}$ is dominant and the increase in local temperature $T_{{\rm PC}}$ occurs. In this model, the relation $T_{{\rm PC}}^{2}=T_{{\rm bath}}^{2}+V^{2}/4L$ is satisfied, where $T_{{\rm bath}}$ and $L$ are the bath temperature and the Lorenz number, respectively\cite{11}. This leads to an approximate expression $T_{{\rm PC}}=3.2$(K/mV)$\times V$(mV), when $T_{{\rm PC}}\gg T_{{\rm bath}}$. Since the largest bias voltage 2mV in figure~\ref{fig2} corresponds to $T_{{\rm PC}}=6.4$ K, the temperature dependence of $\rho$ in figure~\ref{fig1} suggests that the spectra should be V-shaped in the thermal model.

Although some of the spectra are nearly V-shaped in figure~\ref{fig2} (b), the broad maximum at about $\pm$ 1 mV observed for Contacts C and D cannot be explained by the thermal model. Moreover, such a double maximum structure, which resembles the spectra observed in many superconductors\cite{20}, cannot be ascribed to superconductivity, since it does not disappear well above $T_{{\rm c}}$ as described below. The similarity of the voltage width between the zero-bias peak in figure~\ref{fig2} (a) and the double maximum structure in figure~\ref{fig2} (b) may suggest that the spectra for Contacts C and D also reflect $R_{{\rm SHA}}$; the difference in the interface barrier determines whether a peak or a dip appears at zero bias, as observed for superconductors in the BTK theory\cite{21}.

As seen in figures~\ref{fig2} and ~\ref{fig3}, some of the contacts have shown an asymmetry of ${\rm d}V/{\rm d}I$ curves versus $V$: ${\rm d}V/{\rm d}I(+|V|)<{\rm d}V/{\rm d}I(-|V|)$. If this asymmetry has a spectroscopic origin, it will reflect energy-dependent electronic DOS (density of states) around the Fermi level. However, the fact that the asymmetry is remarkable for the contacts following the thermal model in figure~\ref{fig2} (b) and at higher bias voltages in figure~\ref{fig3} suggests that the thermoelectric voltage caused by the increase in temperature at the point contacts is a more likely origin\cite{11}, although the details of the Seebeck coefficient of CePt$_{3}$Si is still not clear. 
\begin{figure}
\begin{center}
\includegraphics*[width=0.9\linewidth, trim=0.5cm 6cm 0.5cm 14cm, clip]{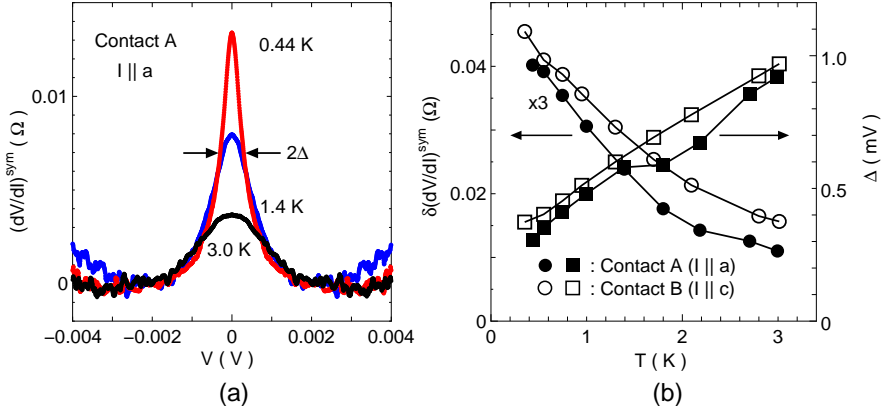}
\caption{\label{fig4}(a) Symmetric part of ${\rm d}V/{\rm d}I$ spectra for Contact A, where $({\rm d}V/{\rm d}I)^{{\rm sym}}$ at $V$=2 mV is subtracted. (b) Temperature dependence of peak height $\delta({\rm d}V/{\rm d}I)^{{\rm sym}}$ and $\Delta$ derived from the symmetric part of ${\rm d}V/{\rm d}I$ spectra for Contact A ($I \parallel a$) and Contact B ($I \parallel c$) in figure~\ref{fig2}(a).}
\end{center}
\end{figure}
%

Hereafter, we focus on the zero-bias peak that is common to HFS with a magnetic order. The evolution of the peak structure is examined by extracting the symmetric part from the spectra using the equation
\begin{equation}
\left(\frac{{\rm d}V}{{\rm d}I}\right)^{{\rm sym}}=\frac{1}{2}\left[\frac{{\rm d}V}{{\rm d}I}(V)+\frac{{\rm d}V}{{\rm d}I}(-V)\right].
\end{equation}
Typical examples of the symmetric part are shown in figure~\ref{fig4}(a) for the spectra in figure~\ref{fig3}. The peak height $\delta({\rm d}V/{\rm d}I)^{{\rm sym}}$ is defined as $\delta({\rm d}V/{\rm d}I)^{{\rm sym}}=({\rm d}V/{\rm d}I)^{{\rm sym}}_{V={\rm 0mV}}-({\rm d}V/{\rm d}I)^{{\rm sym}}_{V={\rm 2mV}}$. The half-width $2\Delta$ of the peak is defined at half-height. The temperature dependence of both properties are given in figure~\ref{fig4}(b) for Contact A ($I\parallel a$) and Contact B ($I\parallel c$) in figure~\ref{fig2} (a). For both contacts, $\Delta$ decreases with a decrease in temperature. This is in contrast to the behaviour observed for URu$_{2}$Si$_{2}$; $\Delta$ appears at $T_{{\rm N}}$, slightly increases by lowering the temperature, which is regarded as the growth of the energy gap at the Fermi surface\cite{11}.  

\begin{table}
\caption{Properties of heavy-fermion superconductors coexisting with a magnetic order. For the U-based materials, the data in \cite{12} are tabulated. }
\label{table1}
\begin{indented}
\item[]\begin{tabular}{cccc}
\br
&$T_{{\rm N}}$(K)&$\Delta$(meV)&$\Delta_{{\rm cr}}$(meV)\\
\mr
URu$_{2}$Si$_{2}$&17.5&9.5&9.9\\
UPd$_{2}$Al$_{3}$&14&13&13\\
UNi$_{2}$Al$_{3}$&4.6&10&\\
CePt$_{3}$Si&2.2&$>0.3$&1\cite{8}\\
\br
\end{tabular}
\end{indented}
\end{table}

Table~\ref{table1} summarizes the characteristics of heavy-fermion superconductors that show a zero-bias peak. For the U-based materials, two possible interpretations of the peak structure has been proposed. One is a gap opening on the Fermi surface due to spin-density-wave formation in the antiferromagnetic state\cite{11,12}. The other interpretation is based on the fact that $\Delta$ is comparable to the crystal field splitting $\Delta_{{\rm cr}}$ between the two levels that are connected by a non-zero matrix element; $\Delta$ reflects a spin-wave gap determined by $\Delta_{{\rm cr}}$.\cite{12}

Although the temperature dependence of the peak for CePt$_{3}$Si is different from URu$_{2}$Si$_{2}$ as described above, $\Delta$ in figure~\ref{fig4} (b) is an order of magnitude smaller than $\Delta$ for the other U-based materials, suggesting the possible scaling law between $\Delta$ and $T_{{\rm N}}$. In addition, $\Delta_{{\rm cr}}$ is comparable to $\Delta$. It should be noted that the spin-wave gaps derived from heat capacity and resistivity data also have similar values: 2.7 K (0.23 mV) and 1.8 K (0.16 mV), respectively\cite{2}. Although $\Delta$ is of the same order of magnitude as various physical properties related to antiferromagnetism in CePt$_{3}$Si, the lack of a drastic change of the spectra at $T_{{\rm N}}$ raises a question about its antiferromagnetic origin. A comprehensive interpretation of the zero-bias peak is left to future study.

As for the spectroscopic evidence of superconductivity, we could not observe it in contrast to the clear unconventional spectra in the normal state, as shown in figure~\ref{fig3}. The absence of a superconducting anomaly has also been reported for URu$_{2}$Si$_{2}$, especially in high-resistance contacts\cite{22,23}, and ascribed to the normal-conducting layer near the surface of URu$_{2}$Si$_{2}$. One possible origin of such a layer is the deformation in the contact region and it probably occurs in CePt$_{3}$Si as well. Even if such a deformed region becomes superconducting, it will only show a reduced anomaly, which is often characterized by the finite quasiparticle lifetime in the BTK theory\cite{11}. Moreover, the unconventional spectra in the normal state of CePt$_{3}$Si that grow at lower temperatures and on a lower voltage scale than in URu$_{2}$Si$_{2}$ ($T<17.5$ K, $V<10$ mV) tend to hide the superconducting anomaly on the scale of the energy gap ($<1$ mV), so that the superconducting anomaly has hardly been observed.

\begin{figure}
\begin{center}
\includegraphics*[width=0.8\linewidth, trim=3cm 3cm 1cm 8cm, clip]{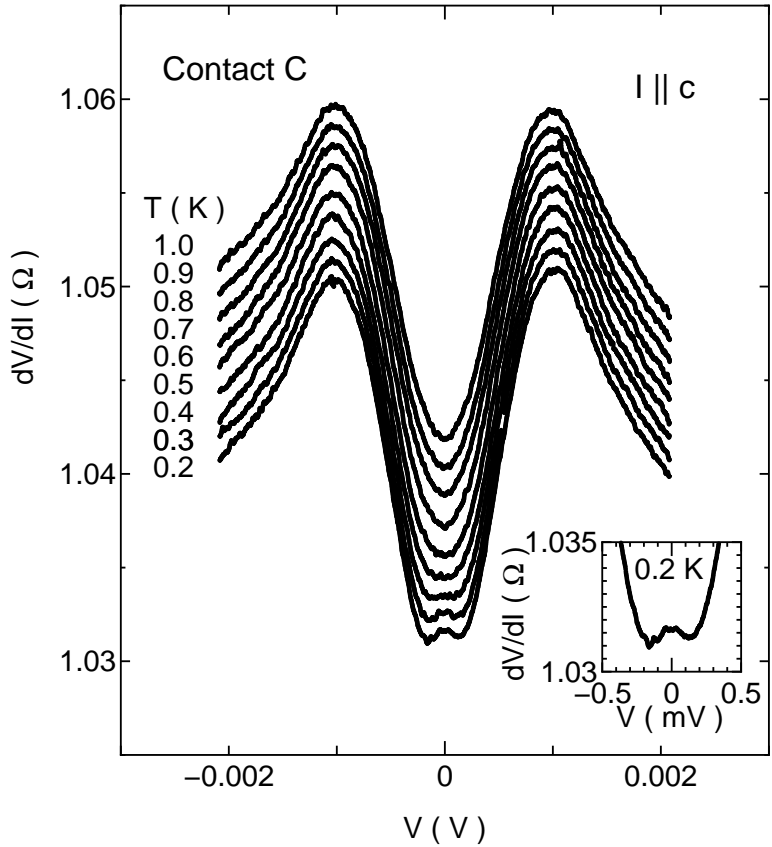}
\caption{\label{fig5}Temperature dependence of ${\rm d}V/{\rm d}I$ versus $V$ characteristics for Contact C. The spectra are shifted vertically by 0.001 $\Omega$ for clarity. Inset shows the superconducting gap at 0.2 K.}
\end{center}
\end{figure}
%
The only exception is Contact C in figure~\ref{fig2} (b). Figure~\ref{fig5} shows the temperature dependence of the spectra for Contact C. The double maximum structures at about $\pm 1$ mV change little within this temperature range. In addition, a small local maximum appears at zero bias below $T_{{\rm c}}\sim$0.6 K. Such a structure is expected when a superconductor and a normal metal are separated by an interface barrier; the voltage $|V_{{\rm S}}|\sim0.15$ mV where the double minimum appears indicates the superconducting energy gap\cite{21}. This value, however, only gives the lower limit of the gap, since the double maximum structure outside is so steep that it tends to make the minimum appear at lower values. Still, $|eV_{{\rm S}}|\sim0.15$ meV is larger than the weak-coupling limit $1.75k_{{\rm B}}T_{{\rm c}}$ = 0.09 meV with $T_{{\rm c}}\sim$ 0.6 K, indicating the strong coupling superconductivity in CePt$_{3}$Si. 

In conclusion, point-contact spectroscopy of CePt$_{3}$Si has shown a dip or a peak of the differential resistance at zero bias. The dip structure varies from contact to contact, while the peak has a half-width of the order of the crystal field splitting or the spin-wave gap. Such a peak structure is commonly observed for heavy-fermion superconductors that coexist with a magnetic order, although for CePt$_{3}$Si the peak disappears not at $T_{{\rm N}}$ but well above $T_{{\rm N}}$.  

\ack
This study was partly supported by a Grant-in-Aid from the Ministry of Education, Culture, Sports, Science and Technology (MEXT), Japan. One of us (Y. \={O}.) was financially supported by the Grant-in-Aid for COE Research (10CE2004) of the MEXT, Japan.

\end{document}